\documentstyle[12pt]{article}
\newcommand{\ba}{\begin{array}}
\newcommand{\ea}{\end{array}}
\newcommand{\beg}{\begin{eqnarray}}
\newcommand{\eeq}{\end{eqnarray}}
\newcommand{\bg}{\begin{eqnarray*}}

\newcommand{\ed}{\end{eqnarray*}}

\newcommand{\nn}{\nonumber}

\newcommand{\p}{\partial}

\newcommand{\notlhd}{\lhd\kern-.8em{/}\ }
\newcommand{\notexist}{\ \exists\kern-.5em{\raise.1em\hbox{/}}\ }

\newcommand{\no}{\noindent}

\newcommand{\pde}[2]{\frac{\p #1}{\p #2}}
\newcommand{\ppd}[3]{\frac{\p^2 #1}{\p #2\p #3}}
\newcommand{\pdd}[2]{\frac{\p^2 #1}{\p #2^2}}
\newcommand{\inp}{{\mbox{\vbox{\hrule width0ex\hbox{\vrule
 height0ex\kern3.8pt
\vbox{\kern2.5pt}\kern3.8pt \vrule height1.6ex}
\hrule width1.6ex}}}}
\newcommand{\bm}[1]{\mbox{\boldmath$#1$}}
\begin{document}

\begin{center}
\no {\Large{\bf $n$-Dimensional Bateman Equation}}\\
\no {\ \ }\\
\no{\bf and }\\
\no{\ \ }\\
\no{\Large{\bf Painlev\'e Analysis
of Wave Equations }}\\
\no{\ \ }

\vspace{0.4cm}
\large

by

\no {\ \ }\\

\vspace{0.4cm}

Norbert Euler$^1$ and Ove Lindblom$^2$
%{\footnote{Norbert@sm.luth.se}} and
%Ove Lindblom{\footnote{Ove@sm.luth.se}}

\vspace{0.5cm}

Department of Mathematics\\
Lule\aa \ University of Technology\\
S-971 87 Lule\aa,\ Sweden\\
\ \ \\
E-mails: $^1$Norbert@sm.luth.se, $^2$Ove@sm.luth.se
\end{center}

\baselineskip 0.6cm

\vspace{0.5cm}

\noindent
{\bf Abstract:} In the Painlev\'e analysis of nonintegrable partial
differential equations one obtains differential constraints describing the
movable singularity manifold. We show that, for a class of $n$-dimensional
wave equations,
these constraints have a general structure which is related to the
$n$-dimensional Bateman equation. In
particular,
we
derive the expressions of the singularity manifold constraint
for the $n$-dimensional sine-Gordon -, Liouville -,
Mikhailov -, and double sine-Gordon equation, as well as two
2-dimensional polynomial field theory equations, and prove that their
singularity manifold
conditions are satisfied by the
$n$-dimensional Bateman equation.
Finally we give some examples.

%\vspace{1cm}

%\noindent
%{\bf Subject Classification:} 35A30, 35C99, 35Q99

%\vspace{1cm}

%\noindent
%{\bf Key words:} multidimensional
%nonlinear partial differential equations,
%integrability, Painlev\'e analysis.

\pagebreak

\baselineskip 0.6cm

%\begin{center}
\section{Introduction}
%\end{center}

\setcounter{equation}{0}

\noindent
The Painlev\'e analysis, as a test for integrability
of partial differential equations (PDEs), was proposed by
Weiss, Tabor and Carnevale in 1983 [26].
It is a generalization of the singular point analysis for
ordinary differential equations (ODEs),
which dates back to the work of Sofia
Kovalevskaya of 1889 [11].
She studied the Euler-Poisson equations
in the complex domain and found conditions under which the only
movable singularities exhibited by the solutions were ordinary poles,
leading to her discovery of a new first integral.
In the late ninteenth century Paul Painlev\'e
completely classified first order ODEs [17], as well as a large class
of second order ODEs [18, 19], on the basis that the only
movable
singularities their solutions exibit,
are ordinary poles. This special property is today known as the
the Painlev\'e property (see, for example [4, 12, 20]).
We also say that an ODE is of {\it Painlev\'e
type},
by which we mean that it belongs to the class of equations in
Painlev\'e's classification, or that it can be transformed to
one of the equations in that class; therefore an ODE which
has the Painlev\'e property. The list of ODEs, classified by
Painlev\'e, is given in the book of Davis [5].

\strut\hfill

%The difference between analytic functions of one variable (in ODEs)
%and analytic functions in several variables (in PDEs) is that the
%singularities of the several variable analytical functions are not
%isolated. If $f=f(x_0,x_1,\ldots,x_{n-1})$ is an analytic function of
%$n$ complex variables $x_0,x_1,\ldots,x_{n-1}$, then the singularities of
%$f$ are in manifolds of $(2n-2)$ dimensions. These manifolds are determined
%by the condition $\phi(x_0,x_1,\ldots,x_{n-1})=0$, where $\phi$ is an
%analytic function in the neighbourhood of the manifold define by it. When
%this manifold depends on the initial conditions, it is called a movable
%singularity manifold.

We consider a PDE to be integrable if it can be solved by an
inverse scattering transform (we refer to the book [1],
and references theirin).
A PDE which is integrable possess the Painlev\'e property,
which means that its
solutions are single-valued in the neighbourhood of non-characteristic
movable singularity manifolds [1, 15, 21].
In this sense the method described by
Weiss, Tabor and Carnevale [26] proposes a
necessary
condition of integrability, also known as the Painlev\'e test,
which is analogous to the algorithm for ODEs
described  by Ablowitz, Ramani and Segur [2] which determines
whether a
given ODE has the Painlev\'e property. One seeks a solution of a given
PDE (in rational form) in the form of a Laurent series (also known as
the Painlev\'e expansion)
\beg
\label{genpain}
u(\bm x)=\phi^{-m}(\bm x)\sum_{j=0}^\infty
u_j(\bm x)\phi^j(\bm x),
\eeq
where $u_j(\bm x)$ are analytic functions of the complex variables
$\bm x=(x_0,x_1,\ldots,x_{n-1})$ (we do not change notation
for the complex domain), with $u_0\neq 0$, in the neighbourhood of a
non-characteristic movable singularity manifold defined by $\phi(\bm
x)=0$ (the pole manifold), where $\phi$ is an analytic function
of $\bm x$. The PDE is said to pass the
Painlev\'e test if, on
substituting (\ref{genpain}) in the PDE, one obtains the correct
number of arbitrary functions as required by the Cauchy-Kovalevsky
theorem, given by the expansion coefficients in (\ref{genpain}),
whereby $\phi$ should be one of the arbitrary functions. The coefficient
in the Painlev\'e expansion, where the arbitrary functions are to
appear, are known as the resonances. If a PDE satisfies the Painlev\'e
test, it is usually [16]
possible to construct B\"acklund transformations
and Lax pairs [6, 20, 24],
which then proves the sufficient condition of integrability.

\strut\hfill

Recently some attention was given to the construction of
exact solutions of nonintegrable  PDEs by the use of a truncated
Painlev\'e series [3, 7, 22, 23].
On applying the Painlev\'e expansion to nonintegrable
PDEs one obtains conditions on $\phi$ at the resonances; the
singular manifold conditions. By truncating the series one usually
obtains
additional constraints on the singularity manifolds, leading to
compatibility problems for the solution of $\phi$
[7, 23, 25]. It has been
known for some time that the 2-dimensional Bateman equation
\beg
\label{phi}
\phi_{x_0x_0}\phi_{x_1}^2+\phi_{x_1x_1}\phi_{x_0}^2-
2\phi_{x_0}\phi_{x_1}\phi_{x_0x_1}=0,
\eeq
plays an important role in the Painlev\'e analysis of
2-dimensional nonintegrable PDEs [25].

\strut\hfill

In the present
paper we show that the general solution of the $n$-dimensional
Bateman equation,
as generalized by Fairlie [9], solves the singularity manifold
condition at the resonance for a class of wave equations. In the
present
paper we
consider the
$n$-dimensional ($n\geq 3$) sine-Gordon -, Liouville -, Mikhailov
equation, and double sine-Gordon equation.
The Painlev\'e test of the
2-dimensional double sine-Gordon equation was analyzed by Weiss [25],
and
resulted in the singularity constrained (\ref{phi}).
Weiss pointed out that the 2-dimensional Bateman equation (\ref{phi})
can be linearized by a Legendre transformation. Moreover, it
is invariant under the Moebius group and admits the general
implicit
solution
\beg
\label{sss}
x_0f_0(\phi)+x_1f_1(\phi)=c,
\eeq
where $f_0$ and $f_1$ are arbitrary smooth functions
and $c$ is an arbitrary real constant.
In the following section we derive the explicit relation between the
singularity manifold
and the 2-dimensional Bateman equation for two
2-dimensional polynomial wave equations. Finally we give some examples
which demonstrate the use of our Propositions.

%\begin{center}
\section{Propositions}
%\end{center}

\setcounter{equation}{0}

\noindent
Fairlie [9] proposed the
following $n$-dimensions Bateman equation:
\beg
\label{genphi}
det\left(\ba{ccccc}
0&\phi_{x_0}&\phi_{x_1}&\cdots &\phi_{x_{n-1}}\\
\phi_{x_0}&\phi_{x_0x_0}&\phi_{x_0x_1}&\cdots &\phi_{x_0x_{n-1}}\\
\phi_{x_1}&\phi_{x_0x_1}&\phi_{x_1x_1}&\cdots &\phi_{x_1x_{n-1}}\\
\vdots&\vdots&\vdots&\vdots&\vdots\\
\phi_{x_{n-1}}&\phi_{x_0x_{n-1}}&\phi_{x_1x_{n-1}}&\cdots &
\phi_{x_{n-1}x_{n-1}}
\ea
\right)=0.
\eeq
Equation (\ref{genphi}) generalizes the 2-dimensional Bateman
equations (\ref{phi}) in $n$ dimensions. It
admits the
following general implicit solution [9]
\beg
\label{gensss}
\sum_{j=0}^{n-1}x_j\,f_j(\phi)=c,
\eeq
where $f_j$ are $n$ arbitrary smooth functions.

\strut\hfill

We consider the $n$-dimensional generalization of
the well known 2-dimensional
sine-Gordon -, Liouville -, and Mikhailov equations,
given respectively by
\beg
\label{set1}
& & \Box_n u+\sin u=0\nn\\
& & \Box_n u+\exp(u)=0\\
& & \Box_n u+\exp(u)+\exp(-2u)=0\nn,
\eeq
as well as the
double
sine-Gordon equation in $n$ dimensions:
\beg
\label{dsgn}
\Box_n u+\sin\frac{u}{2}+\sin u=0.
\eeq
Here $\Box_n$ denotes the d'Alembert operator in
$n$-dimensional Minkowski
space, and is defined by
\bg
\Box_n:=\pdd{\ }{x_0}-\sum_{j=1}^{n-1}\pdd{\ }{x_j}.
\ed
It is well known that the wave equations (\ref{set1}) are integrable
for $n=2$ (see, for example, [1]).

\strut\hfill

Before we state our Proposition
for the singularity manifolds of those equations, we
introduce some notations
and a Lemma. We call the  $(n+1)\times (n+1)$-matrix,
the determinant of which defines
the $n$-dimensional Bateman equation (\ref{genphi}),
the $n$-dimensional Bateman matrix and denote
this matrix by $B_{n+1}^n$. The subscript of $B$ shows the
size of the matrix while the superscript gives the dimension
(the number of variables of $\phi$), i.e.,
for the $n$-dimensional Bateman matrix (\ref{genphi}), the
associated Bateman matrix is
\beg
\label{bphi}
B_{n+1}^n=\left(\ba{ccccc}
0&\phi_{x_0}&\phi_{x_1}&\cdots &\phi_{x_{n-1}}\\
\phi_{x_0}&\phi_{x_0x_0}&\phi_{x_0x_1}&\cdots &\phi_{x_0x_{n-1}}\\
\phi_{x_1}&\phi_{x_0x_1}&\phi_{x_1x_1}&\cdots &\phi_{x_1x_{n-1}}\\
\vdots&\vdots&\vdots&\vdots&\vdots\\
\phi_{x_{n-1}}&\phi_{x_0x_{n-1}}&\phi_{x_1x_{n-1}}&\cdots &
\phi_{x_{n-1}x_{n-1}}
\ea
\right).
\eeq
In particular the submatrices of the above $n$-dimensional
Bateman matrix are of importance, i.e., the submatrices
$B_p^n$, where $ 3\leq p\leq n+1$. These submatrices, which we call
$n$-dimensional Bateman
submatrices, are obtained by deleting
rows and corresponding columns of $B_{n+1}^n$. We give the following

\strut\hfill

\noindent
{\sc DEFINITION.} {\it Let
\bg
M_{x_{j_1}x_{j_2}\ldots x_{j_r}}
\ed
denote the determinant of a Bateman submatrix, that remains
after the rows
and columns containing the derivatives $\phi_{x_{j_1}}$,
$\phi_{x_{j_2}}$, \ldots, $\phi_{x_{j_r}}$ have been deleted
from the $n$-dimensional Bateman matrix (\ref{bphi}). Let
\bg
j_1,\ldots,j_r\in\{0,1,\ldots,n-1\},\quad j_1<j_2<\cdots < j_r,\quad
r\leq n-2,\quad {\mbox{for}}\
n\geq 3.
\ed
Then ${\displaystyle M_{x_{j_1}x_{j_2}\ldots x_{j_r}}}$ are the
determinants of the
Bateman matrices
$B_{n+1-r}^n$. We call the equations
\beg
M_{x_{j_1}x_{j_2}\ldots x_{j_r}}=0
\eeq
the minor $n$-dimensional Bateman equations.}

\strut\hfill

Note that the $n$-dimensional Bateman equation (\ref{genphi})
has
$
n!/[r!(n-r)!]
$
minor $n$-dimensional Bateman equations. Consider an example:
If $n=5$ and $r=2$,
then there exist 10 minor Bateman equations, one of which is
given by $M_{x_2x_3}$, i.e.,
\beg
det\left(\ba{cccc}
0&\phi_{x_0}&\phi_{x_1}&\phi_{x_4}\\
\phi_{x_0}&\phi_{x_0x_0}&\phi_{x_0x_1}&\phi_{x_0x_4}\\
\phi_{x_1}&\phi_{x_0x_1}&\phi_{x_1x_1}&\phi_{x_1x_4}\\
\phi_{x_4}&\phi_{x_0x_4}&\phi_{x_1x_4}&\phi_{x_4x_4}
\ea
\right)=0.
\eeq

\strut\hfill

We can now state the following

\strut\hfill

\noindent
{\sc LEMMA.}  {\it If $\phi$ satisfies the $n$-dimensional Bateman equation
(\ref{genphi}), then it satisfies any minor
Bateman equation
\bg
M_{x_{j_1}x_{j_2}\ldots x_{j_r}}=0
\ed
with
\bg
j_1,\ldots,j_r\in\{0,1,\ldots,n-1\},\quad j_1<j_2<\cdots < j_r,\quad
r\leq n-2,\quad {\mbox{for}}\
n\geq 3.
\ed
}

\strut\hfill

{\it Proof:}
By implicitly differentiating the general solution (\ref{gensss}) of
the $n$-dimensional Bateman equation (\ref{genphi}), it is easily shown
that any minor $n$-dimensional Bateman equation is satisfies
by this solution. Since (\ref{gensss}) is the general solution of the
$n$-dimensional Bateman equation, the
proof is concluded.    \hfill$\qquad\Box$

\strut\hfill

We now prove

\strut\hfill

\noindent
{\sc PROPOSITION 1.} {\it For $n\geq 3$, the singularity manifold
conditions of the
$n$-dimensional
sine-Gordon -, Liouville -, and Mikhailov equations
(\ref{set1}), are satisfied by
the solution of the $n$-dimensional Bateman equation (\ref{genphi}).}

\strut\hfill

{\it Proof:} We do the
proof for the sine-Gordon equation. For the
Liouville - and Mikhailov equation, the proofs are similar.
By the substitution
\bg
v(\bm x)=\exp[iu(\bm x)]
\ed
the $n$-dimensional sine-Gordon equation takes the following form:
\beg
\label{nsg}
v\Box_n
v-\left(\bigtriangledown_n v\right)^2+\frac{1}{2}\left(v^3-v\right)=0,
\eeq
where
\bg
\left(\bigtriangledown_n v\right)^2:=\left(\pde{v}{x_0}\right)^2-
\sum_{j=1}^{n-1}\left(\pde{v}{x_j}\right)^2.
\ed
The dominant behaviour of (\ref{nsg}) is 2, so that the Painlev\'e
expansion is
\bg
v(\bm x)=\sum_{j=0}^\infty v_j(\bm x)\phi^{j-2}(\bm x).
\ed
The resonance is at 2 and the first two coefficients
in the Painlev\'e expansion have the following form:
\bg
v_0=-4\left(\bigtriangledown_n \phi\right)^2, \qquad v_1=4\Box_n \phi.
\ed
We first consider $n=3$. The singularity manifold condition at the
resonance is then given by
\bg
det\left(\ba{cccc}
0&\phi_{x_0}&\phi_{x_1}&\phi_{x_{2}}\\
\phi_{x_0}&\phi_{x_0x_0}&\phi_{x_0x_1}&\phi_{x_0x_{2}}\\
\phi_{x_1}&\phi_{x_0x_1}&\phi_{x_1x_1}&\phi_{x_1x_{2}}\\
\phi_{x_{2}}&\phi_{x_0x_{2}}&\phi_{x_1x_{2}}&
\phi_{x_{2}x_{2}}
\ea
\right)=0,
\ed
which is the 3-dimensional Bateman equation $\det B_4^3=0$, as defined by
(\ref{genphi}).

\strut\hfill

Consider now $n\geq 4$. The condition at the
resonance can be written as follows:
\beg
\label{thpn}
\sum_{j_1,j_2,\ldots, j_{n-3}=1}^{n-1}
M_{ x_{j_1}x_{j_2}\ldots x_{j_{n-3}} }-\sum_{j_1,j_2,\ldots, j_{n-4}=1}^{n-1}
M_{x_0x_{j_1}x_{j_2}\ldots
x_{j_{n-4}}}=0,
\eeq
where
\bg
j_1<j_2<\cdots < j_{n-3},
\ed
and $\displaystyle{ M_{ x_{j_1}x_{j_2}\ldots x_{j_{n-3}} } }$,
$\displaystyle{M_{x_0x_{j_1}x_{j_2}\ldots x_{j_{n-4}}} }$
are minor $n$-dimensional Bateman equations, i.e.,
the determinants of $4\times 4$ Bateman matrices $B_4^n$.
By the Lemma give above, equation (\ref{thpn})  is
satisfied by the solution of the $n$-dimensional Bateman equation
(\ref{genphi}).\hfill
{$\Box$}

\strut\hfill

We now consider the double
sine-Gordon equation in $n$ dimensions (\ref{dsgn}):
\bg
%\label{dsgn}
\Box_n u+\sin\frac{u}{2}+\sin u=0.
\ed
It was shown by Weiss [25], that for $n=2$ this equation does not pass the
Painlev\'e test, and that the singularity manifold condition is
given by the Bateman equation (\ref{phi}).

For $n$ dimensions we
prove the following

\strut\hfill

\noindent
{\sc PROPOSITION 2.} {\it For $n\geq 2$, the singularity manifold
condition of the
double sine-Gordon equation (\ref{dsgn})
is satisfied by the solution of
the $n$-dimensional Bateman equation (\ref{genphi}).}

\strut\hfill

{\it Proof:} By the substitution
\bg
v(\bm x)=\exp\left[\frac{i}{2}\,u(\bm x)\right]
\ed
the rational form of the double sine-Gordon equation (\ref{dsgn})
is obtained as
\bg
v\Box v+(\bigtriangledown v_n)^2+\frac{1}{4}(v^3-v)+\frac{1}{4}(v^4-1)=0.
\ed
The Painlev\'e expansion takes the form
\bg
v(\bm x)=\sum_{j=0}^\infty v_j(\bm x)\phi^{j-1}(\bm x)
\ed
and the resonance is 2. The first two expansion coefficients are
\bg
v_0=-4(\bigtriangledown_n v)^2,\qquad
v_1=\frac{2}{v_0}\Box_n\phi-\frac{1}{2}
\ed
For the singularity manifold condition we have to consider four cases:

\strut\hfill

\noindent
{\bf Case $n=2$:} At the resonance we obtain (\ref{phi}), i.e.,
%the
%determinant of the $3\times 3$ Bateman matrix in $x_0$ and $x_1$,
%equal to zero. Thus
\bg
\det B_{3}^2=0.
\ed

\noindent
{\bf Case $n=3$:} The condition now takes the following form:
\bg
8 \det B_4^3+\left(M_{x_1}+M_{x_2}-M_{x_0}\right)v_0=0.
\ed
%{\bf Case $n=4$:} The condition is
%\bg
%8\left(M_{x_1}+M_{x_2}+M_{x_3}-M_{x_0}\right)
%+\left(M_{x_1x_2}+M_{x_1x_3}+M_{x_2x_3}-M_{x_0x_1}-M_{x_0x_2}-M_{x_0x_3}
%\right)v_0=0.
%\ed

\noindent
{\bf Case $n\geq 4$:} The condition at the resonance can be written as
follows:
\bg
& & 8\left(\sum_{j_1,j_2,\ldots j_{n-3}=1}^{n-1}M_{x_{j_1}x_{j_2}\ldots
x_{j_{n-3}}}\right)\\
& & \qquad\qquad+\left(\sum_{j_1,j_2,\ldots j_{n-2}=1}^{n-1}
M_{x_{j_1}x_{j_2}\ldots
x_{j_{n-2}}}-\sum_{j_1,j_2,\ldots j_{n-3}=1}^{n-1}
M_{x_0x_{j_1}x_{j_2}\ldots
x_{j_{n-3}}}\right)v_0=0,
\ed
where
\bg
j_1<j_2<\cdots <j_{n-3}<j_{n-2}.
\ed

\strut\hfill

By the above Lemma the proof is concluded.\hfill $\Box$

\strut\hfill

We now consider two well known nonlinear polynomial field theory
equations,
the
so-called
nonlinear Klein-Gordon equations:
\beg
\label{kg1}
\Box_2 u+u^k=0
\eeq
with $k=2,\  3$.
In light-cone coordinates, i.e.,
\bg
x_0\longrightarrow \frac{1}{2}(x_0-x_1),\qquad
x_1\longrightarrow \frac{1}{2}(x_0+x_1),
\ed
(\ref{kg1}) takes the form
\beg
\label{kg2}
\ppd{u}{x_0}{x_1}+u^k=0.
\eeq
It should be noted that the 2-dimensional Bateman equation
remains invariant under the light-cone coordinates. Therefore,
for our purpose we can work with (\ref{kg2}) instead of (\ref{kg1}).
In [8] it was shown that the nonlinear Klein-Gordon equation
(\ref{kg2}), with $k=3$, does not pass the Painlev\'e test. We are now
interested
in the relation between the 2-dimensional Bateman equation
(\ref{phi}) and the singularity
manifold condition of (\ref{kg2}) for the case $k=2$ as well as $k=3$.

\strut\hfill

We prove the following

\strut\hfill

\noindent
{\sc PROPOSITION 3.} {\it The solution of the 2-dimensional Bateman
equation (\ref{phi}) satisfies the singularity manifold condition
of the nonlinear Klein-Gordon equation (\ref{kg2}) for $k=2$ and
$k=3$.}

\strut\hfill

{\it Proof:} First we
consider equation (\ref{kg2}) with $k=3$, i.e.,
\beg
\label{kg3}
\ppd{u}{x_0}{x_1}+u^3=0.
\eeq
For the Painlev\'e expansion
\beg
\label{genpain3}
u(x_0, x_1)=\phi^{-m}(x_0,x_1)\sum_{j=0}^\infty
u_j(x_0,x_1)\phi^j(x_0,x_1),
\eeq
we find
that the dominant behaviour is
-1, the resonance is 4, and the first three expansion coefficients
in expansion (\ref{genpain3}) are
\bg
& & u_0^2=2\phi_{x_0}\phi_{x_1},\\
& & u_1=-\frac{1}{3u_0^2}\left(u_0\phi_{x_0x_1}+u_{0x_1}\phi_{x_0}
+u_{0x_0}\phi_{x_1}\right),\\
& & u_2=\frac{1}{3u_0^2}\left(u_{0x_0x_1}-3u_0u_1^2\right),\\
& & u_3=\frac{1}{u_0^2}\left(u_2\phi_{x_0x_1}+u_{2x_1}\phi_{x_0}
+u_{2x_0}\phi_{x_1}+u_{1x_0x_1}-6u_0u_1u_2\right).
\ed
At the resonance we obtain the following singularity manifold
condition:
\beg
\label{singman2}
\Phi\sigma-\left(\phi_{x_0}\Phi_{x_1}-\phi_{x_1}\Phi_{x_0}\right)^2=0,
\eeq
where $\Phi$ is the 2-dimensional Bateman equation
given by (\ref{phi}), i.e.,
\bg
\Phi=
\phi_{x_0x_0}\phi_{x_1}^2+\phi_{x_1x_1}\phi_{x_0}^2-
2\phi_{x_0}\phi_{x_1}\phi_{x_0x_1},
\ed
and
$\sigma$ contains derivatives of $\phi$ with respect to
$x_0$ and $x_1$. The explicit form of $\sigma$ is not interesting for
our proof.
The explicit appearance of $\Phi$ (\ref{singman2}) concludes the proof
for the nonlinearity $k=3$.

\strut\hfill

For the equation
\beg
\label{pf2}
\frac{\p^2 u}{\p x_0\p x_1}+u^2=0
\eeq
the singularity manifold condition is somewhat more complicated.
 The dominant behaviour of (\ref{pf2})
is -2 and the resonance is 6. The first five expansion coefficients
in the Painlev\'e expansion take the following form:
\bg
& & u_0=-6\phi_{x_0}\phi_{x_1},\\
& & u_1=\frac{1}{\phi_{x_0}\phi_{x_1}+u_0}
\left(u_{0x_1}\phi_{x_0}+u_{0x_0}\phi_{x_1}+u_0\phi_{x_0x_1}\right),\\
& & u_2=-\frac{1}{2u_0}\left(u_{0x_0x_1}+u_1^2-u_{1x_1}\phi_{x_0}
-u_{1x_0}\phi_{x_1}-u_1\phi_{x_0x_1}\right),\\
& & u_3=-\frac{1}{2u_0}\left(u_{1x_0x_1}+2u_1u_2\right),\\
& & u_4=-\frac{1}{\phi_{x_1}\phi_{x_0}+u_0}
\left(u_3\phi_{x_0x_1}+u_{2x_0x_1}+2u_1u_3+u_{3x_1}\phi_{x_0}
+u_{3x_0}\phi_{x_1}+u_2^2\right),\\
& & u_5=-\frac{1}{6\phi_{x_0}\phi_{x_1}+2u_0}\left(
2u_1u_4+2u_4\phi_{x_0x_1}+2u_{4x_0}\phi_{x_1}
+2u_{4x_1}\phi_{x_0}+2u_2u_3+u_{3x_0x_1}\right).
\ed
At the resonance the singularity manifold condition is a PDE of
order six, which consists of
372 terms (!) all of which are derivatives of $\phi$ with respect to
$x_0$ and $x_1$. This condition may be written in the following form:
\beg
\label{35}
\sigma_1\Phi+\sigma_2\Psi+\left(
\phi_{x_0}\Psi_{x_1}-\phi_{x_1}\Psi_{x_0}-\sigma_3\Psi-\sigma_4\Phi\right)^2=0,
\eeq
where $\Phi$ is the 2-dimensional Bateman equation
(\ref{phi}), and
\bg
\Psi=\phi_{x_0}\Phi_{x_1}-\phi_{x_1}\Phi_{x_0}, \qquad \Phi=
\phi_{x_0x_0}\phi_{x_1}^2+\phi_{x_1x_1}\phi_{x_0}^2-
2\phi_{x_0}\phi_{x_1}\phi_{x_0x_1}.
\ed
Here $\sigma_1,\ldots,\sigma_4$ consist of derivatives of $\phi$
with respect to $x_0$ and $x_1$. Their explicit form is not interesting.
By (\ref{35}) it is clear that the general solution of the
Bateman equation
satisfies the singularity manifold condition for
(\ref{pf2}).\hfill $\Box$

\strut\hfill

Due to its enormous complexity in higher dimensions, we were not able
to
find the explicit relations
between the
singularity manifold for higher dimensional equations of the form
\beg
\label{nkg}
\Box_n u+u^k=0
\eeq
and the $n$-dimensional Bateman equation (or minor Bateman equations).
We

\strut\hfill

\noindent
{\large CONJECTURE.} {\it 
In $n$-dimensions, the solution of
the $n$-dimensional Bateman equation (\ref{genphi})
satisfies the singularity manifold
condition of (\ref{nkg}) for $k=2, 3$.}

\strut\hfill

Some examples of (\ref{nkg})
are also given below, and these are
consistent with this view.

\strut\hfill

%\begin{center}
\section{Application}
%\end{center}

\setcounter{equation}{0}

\noindent
According to a conjecture by Ablowitz, Ramani and Segur [2], every
ODE that can be obtained by a Lie symmetry reduction (similarity reduction)
of a PDE, which is solvable by the inverse scattering transform method, has
the Painlev\'e property. Some weak form of this conjecture was proved in
[13].
%that if an integrable
%2-dimensional PDE
%is reduced by its Lie symmetry group invariants to ODEs, then
%all reduced equations are of Painlev\'e type, i.e., possess
%the Painlev\'e property for ODE's.
On the other hand, if we would consider a nonintegrable 2-dimensional PDE,
then it is possible that some of the ODEs resulting by
some reduction Ansatz of the PDE, may also be of Painlev\'e type.
In particular, the reduced ODE would fullfil the necessary condition
to be of Painlev\`e type  (pass the Painlev\'e test for ODEs)
for those Ans\"atze for which the new independent
variable satisfies the condition on the singularity manifold of the given PDE.
By the Propositions stated in the previous section, we know that the
condition on the singularity manifold is satisfied by the $n$-dimensional
Bateman equation for our class of equations.
Thus, the Propositions, lead to the following

\strut\hfill

\noindent
{\sc COROLLARY.} {\it
The nonlinear wave equations (\ref{set1}), (\ref{dsgn}),
(\ref{kg3}) (\ref{pf2})
can be reduced to
ODEs which satisfy the necessary condition to be of
Painlev\'e type, if and only if
the new independent variables of the reduced ODEs
satisfy the corresponding $n$-dimensional Bateman equation (\ref{phi}).}

\strut\hfill

This means that if we were to
reduce one of the nonintegrable $n$-dimensional PDEs discussed in our
paper
into an ODE
with independent variable $\omega$ by,
for example, an Ansatz of the form
\beg
\label{ansatz}
u(x_0,x_1,\ldots,x_{n-1})=f_1(x_0,x_1,\ldots,x_{n-1}) \varphi(\omega)
+f_1(x_0,x_1,\ldots,x_{n-1}),
\eeq
then we can easily test the necessary condition of
integrability of the resulting ODE by checking whether
$\omega$ satisfies the $n$-dimensional Bateman equation (\ref{genphi}).
This would be the same as to perform the Painlev\'e test on the resulting
ODE.
By Lie symmetry analysis of PDEs
one is able to systematically construct Ans\"atze which reduce
the PDEs to ODEs according to their Lie transformation group properties
(see for example [10]).
By the above Corollary one is now able to classify the group invariants
(that are independent of $u$)
for the given PDEs, and determine which group invariants may
result in ODE's of
Painlev\'e type, whithout performing the Painlev\'e analysis on the
actual reduced ODEs, but by
merely
checking whether the invariants satisfy the $n$-dimensional
Bateman equation (\ref{genphi}).
One must note that the reduction Ansatz is not necessarily
related to a
classical Lie symmetry invariant. One can obtain very interesting
reduction Ans\"atze by the use of the so-called conditional symmetries, or
$Q$-symmetries (see [10] for some interesting examples).

Below we give some examples of the stated Corollary.
A more systematic
analysis and classification of the the equations treated here,
will be presented in a future paper.

\strut\hfill

\noindent
{\sc EXAMPLE 1.} Consider the 3-dimensional Liouville
equation [10] , i.e.,
\beg
\label{liu3}
\Box_3 u+\lambda \exp(u)=0,
\eeq
with the Ansatz
\beg
\label{ans3}
& & u(x_0,x_1,x_2)=\varphi (\omega)-2
\ln(\alpha_0y_0-\alpha_1y_1-\alpha_2y_2)\nonumber\\
& & \omega(x_0,x_1,x_2)=(\alpha_0y_0-\alpha_1y_1-\alpha_2y_2)
(\beta_0y_0-\beta_1y_1-\beta_2y_2)^a
%(\bm\alpha \cdot \bm y)(\bm\beta\cdot \bm y)^a
\eeq
where $a\in {\cal Q}\backslash \{0\}$ and
\bg
& &\alpha_0^2-\alpha_1^2-\alpha_2^2
=\alpha_0\beta_0-\alpha_1\beta_1-\alpha_2\beta_2=0,\\
& &\beta_0\beta_0-\beta_1\beta_1-\beta_2\beta_2 <0,\\
& & y_\mu=x_\mu+a_\mu,\qquad \mu = 0,1,2.
\ed
Here $\omega$, given by
(\ref{ans3}),
 satisfies the 3-dimensional Bateman equation 
 $\det B_4^3=0$,
so that by the Corollary we are ensured that the reduced ODE,
resulting from Ansatz
(\ref{ans3}),
satisfies the necessary condition to be of
Painlev\'e type.
Ansatz (\ref{ans3}) leads to the following
ODE:
\beg
\label{ode1}
a^2\omega^2
\frac{d^2\varphi}{d\omega^2}+a(a-1)\omega\frac{d\varphi}{d\omega}
+\lambda\exp(\varphi)=0.
\eeq
Equation (\ref{ode1}) is of Painlev\'e type and admits the
general solution
\beg
\label{sol1}
& & \varphi(\omega)=-2\ln\left[
\frac{\sqrt{-\lambda}}{\sqrt{2}c_1}
\omega^{-1/a}\cos(c_1\omega^{1/a}+c_2)\right];\quad
\lambda<0\\
& &\varphi(\omega)=-2\ln\left[\frac{\sqrt{\lambda}}{\sqrt{2}c_1}\omega^{-1/a}
\cosh(c_1\omega^{1/a}+c_2)\right];\quad
\lambda>0.
\eeq
By (\ref{sol1}) and the Ansatz (\ref{ans3}) an exact solution of the
Liouville equation (\ref{liu3}) follows:
\bg
& & \hspace{-3mm}u(x_0,x_1,x_2)=-2\ln\left[\frac{\sqrt{-\lambda}}{\sqrt{2}c_1}\omega^{-1/a}\cos(c_1\omega^{1/a}+c_2)\right]
-2\ln(\alpha_0y_0-\alpha_1y_1-\alpha_2y_2);\ \lambda<0 \\
& &\\
& & \hspace{-3mm}
u(x_0,x_1,x_2)=-2\ln\left[\frac{\sqrt{\lambda}}{\sqrt{2}c_1}\omega^{-1/a}
\cosh(c_1\omega^{1/a}+c_2)\right]
-2\ln(\alpha_0y_0-\alpha_1y_1-\alpha_2y_2);\ \lambda>0\\
& &\\
& &\hspace{-3mm}
\omega(x_0,x_1,x_2)=(\alpha_0y_0-\alpha_1y_1-\alpha_2y_2)
(\beta_0y_0-\beta_1y_1-\beta_2y_2)^a,\quad
y_\mu=x_\mu+a_\mu,\ \mu = 0,1,2.
\ed
This example
can easily be extended to $n$ dimensions.

\strut\hfill

\noindent
{\sc EXAMPLE 2.} Consider the 4-dimensional sine-Gordon equation
[10],
i.e,
\beg
\label{sg4}
\Box_4 u+\sin(u)=0.
\eeq
By the Ansatz
\beg
\label{ans4}
& & u(x_0,x_1,x_2,x_3)=\varphi(\omega)\nonumber\\
& & \omega(x_0,x_1,x_2,x_3)=\frac{x_2-x_3(x_0+x_1)}{\sqrt{1+(x_0+x_1)^2}}+f(x_0+x_1),
\eeq
where $f$ is an arbitrary smooth function of its argument,
(\ref{sg4}) reduces to the following integrable ODE:
\beg
\label{ode2}
\frac{d^2\varphi}{d\omega^2}-\sin\varphi =0.
\eeq
It easy to show that $\omega$, given by (\ref{ans4}),
satisfies the 4-dimensional Bateman equation 
$\det B_5^4=0$.
Equation (\ref{ode2}) can be integrated in terms of Jacobi
elliptic functions to obtain exact solutions of the 4-dimensional
sine-Gordon equation (\ref{sg4}).

\strut\hfill

\noindent
{\sc EXAMPLE 3.} Consider the 2-dimensional nonlinear Klein-Gordon
equation
\beg
\label{kg_2}
u_{x_0x_1}+\lambda u^3=0.
\eeq
We demonstrate that by the given Corollary and the Ansatz
\beg
\label{ans_kg2}
u(x_0,x_1)=h(x_0,x_1)\varphi(\omega),
\eeq
where $\omega$ satisfies the 2-dimensional Bateman equation
(\ref{phi}) i.e.,
\bg
x_0 f_0(\omega)+x_1f_1(\omega)=c,
\ed
we are able to construct ODEs which pass the Painlev\'e test.
Ansatz (\ref{ans_kg2})
leads to
\beg
\label{ode_kg}
& &{\hspace{-6mm}}\left(\frac{fgh}{(x_0\dot f_0+x_1\dot f_1)^2}\right)
\frac{d^2 \varphi}{d \omega^2}\nonumber\\
& & \nonumber\\
& & \qquad+\left(\frac{h(\dot f_0f_1+f_0\dot f_1)}{
(x_0\dot f_0+x_1\dot f_1)^2}
-\frac{fgh(x_0\ddot f_0+x_1\ddot f_1)}{(x_0\dot f_0+x_1\dot
f_1)^3}
-\frac{h_{x_1}f_0+h_{x_0}f_1}{(x_0\dot f_0+x_1\dot f_1)}\right)
\frac{d\varphi}{d\omega}\nonumber\\
& & \nonumber\\
& & \qquad+h_{x_0x_1}\varphi +\lambda h^3\varphi^3=0.
\eeq
Here $h=h(x_0,x_1),\ f_i=f_i(\omega)$, and
$\dot f_i\equiv df_i/d \omega$ $(i=0,1)$.
For example, let
\bg
h(x_0,x_1)=\frac{1}{x_0},\quad f_1(\omega)=-1,
\ed
then (\ref{ode_kg}) reduces to
\beg
\label{gen_pain}
\ddot\varphi+\left(2\frac{\dot f}{f}-\frac{\ddot f}{\dot
f}\right)\dot\varphi -\left(\frac{\lambda\dot f^2}{f}\right)\varphi^3=0.
\eeq
Equation (\ref{gen_pain}) satisfies the necessary condition to be of
Painlev\'e type (it passess the Painlev\'e test for ODEs), which is
in agreement with the above Corollary, as we are using the general
solution of the 2-dimensional Bateman equation (\ref{phi}).
Note that for $f_0(\omega)=\omega$
we obtain the same ODE which was obtained with a Lie symmetry analysis
in [8].
We remark that the use of the general solution (\ref{sss}) of
the Bateman equation (\ref{phi}), in the construction of exact solutions
of (\ref{kg_2}), is clearly limited. A more effective approach, to obtain
exact solutions, would be
to linearize the 2-dimensional Bateman equation by the Legendre
transformation, as outlined by Webb and Zank [23]. However, this is not
the purpose of the present paper.

\strut\hfill

\noindent
{\sc EXAMPLE 4.} Consider the 4-dimensional nonlinear Klein-Gordon
equation
\beg
\label{kg4}
\Box_4 u+\lambda u^3=0,
\eeq
where $\lambda\in {\cal R}$. Assymptotic solutions of
(\ref{kg4}) were constructed in
[14] by the use the Poincar\'e group
$P(1,3)$ and its invariants.
By composing the group invariants, we obatin the following
Ansatz for (\ref{kg4}):
\beg
\label{anskg}
& & \hspace{-5.5mm}u(x_0,x_1,x_2,x_3)=\varphi(\omega)\nonumber\\
& & \hspace{-5.5mm}\omega(x_0,x_1,x_2,x_3)=\beta_1(<\tilde{\bm p},{\bm x}>+a_1)
-\beta_2(<\tilde{\bm \alpha},{\bm x}>+a_2)
-\beta_3(<\tilde{\bm \beta},{\bm x}>+a_3)\\
& &\qquad\qquad +a\ln\left\{\alpha_1
(<\tilde{\bm p},{\bm x}>+a_1)
-\alpha_2(<\tilde{\bm \alpha},{\bm x}>+a_2)
-\alpha_3(<\tilde{\bm \beta},{\bm x}>+a_3)\right\}.\nonumber
\eeq
Here
$<\tilde{\bm p}, {\bm x}>\equiv
p_0x_0-p_1x_1-p_2x_2-p_3x_3,\quad
<\tilde{\bm \alpha}, {\bm x}>\equiv
\alpha_0x_0-\alpha_1x_1-\alpha_2x_2-\alpha_3x_3,$
$<\tilde{\bm \beta}, {\bm x}>\equiv
\beta_0x_0-\beta_1x_1-\beta_2x_2-\beta_3x_3
$
and $a_j$ ($j=0,1,2,3$) are arbitrary real constants, whereas
$\alpha_j, \beta_j,\tilde \alpha_\mu, \tilde \beta_\mu,
\tilde p_\mu$ ($j=1,2,3;\ \mu=0,1,2,3$) are real constants which must
satisfy the following conditions:
\beg
\label{313}
& & \beta_1^2-\beta_2^2-\beta_3^2=-1,\quad
\alpha_1^2-\alpha_2^2-\alpha_3^2=
\alpha_1\beta_1-\alpha_2\beta_2-\alpha_3\beta_3=0\\
& &\nonumber\\
& &
\label{314}
<\tilde{\bm p},\tilde{\bm p}>=1,\quad
<\tilde{\bm \alpha},\tilde{\bm \alpha}>=
<\tilde{\bm \beta},\tilde{\bm \beta}>=-1,\nonumber\\
& &
<\tilde{\bm \alpha},\tilde{\bm \beta}>=
<\tilde{\bm \alpha},\tilde{\bm p}>=
<\tilde{\bm \beta},\tilde{\bm p}>=0.
\eeq

%\tilde p_0^2-\tilde p_1^2-p_2^2-\tilde p_3^2=1,\quad
%\tilde \alpha_0^2-\tilde \alpha_1^2-\tilde\alpha_2^2-\tilde \alpha_3^2=
%\tilde \beta_0^2-\tilde \beta_1^2-\tilde\beta_2^2-\tilde \beta_3^2=-1
%\nonumber\\
%\label{314}
%& &
%\tilde \alpha_0\tilde \beta_0-\tilde \alpha_1\tilde\beta_1
%-\tilde\alpha_2\tilde\beta_2
%-\tilde \alpha_3\tilde\beta_3=
%\tilde \alpha_0\tilde p_0-\tilde \alpha_1\tilde p_1
%-\tilde\alpha_2\tilde p_2
%-\tilde \alpha_3\tilde p_3=0\nonumber\\
%& & 
%\tilde p_0\tilde \beta_0-\tilde p_1\tilde\beta_1
%-\tilde p_2\tilde\beta_2
%-\tilde p_3\tilde\beta_3=0
%\eeq

Here $\omega$, given by (\ref{anskg}), satisfies the 4-dimensional
Bateman equation $\det B_5^4=0$, and the
reduced equation 
\beg
\label{kg4_1}
\frac{d^2\varphi}{d\omega^2}+\lambda \varphi^3=0
\eeq
is of Painlev\'e type.
The general solution of (\ref{kg4_1}) is given in terms of
Jacobi elliptic functions [5].

\strut\hfill

\noindent
{\sc EXAMPLE 5.} Consider the 4-dimensional nonlinear Klein-Gordon equation
\beg
\label{mkg}
\Box_4 u+\lambda_1 u +\lambda_2 u^3=0,
\eeq
where $\lambda_1,\lambda_2\in{\cal R}$. By the invariants of the
Poincar\'e group, and its Lie subalgebras,
the following two Ans\"atze are, for example, possible:
\beg
& & u(x_0,x_1,x_2,x_3)=\varphi(\omega_1)\nonumber\\
& & \omega_1= \frac{c}{2}\left\{<\tilde{\bm \gamma},\bm x>^2+\left[
<\tilde{\bm \beta}, \bm x>+\frac{1}{4}\left(
<\tilde{\bm p}, \bm x>+<\tilde{\bm \alpha},\bm x >\right)^2
\right]^{1/2}\right\}\nonumber\\
& & \phantom{\omega_1}
+q_1<\tilde{\bm \gamma}, \bm x>-
q_2\left[
<\tilde{\bm \beta},\bm x>+\frac{1}{4}\left(<\tilde{\bm p}, \bm x >+
<\tilde{\bm \alpha}, \bm x>\right)^2\right],
\eeq
and
\beg
& & u(x_0,x_1,x_2,x_3)=\varphi(\omega_2)\nonumber\\
& & \omega_2(x_0,x_1,x_2,x_3)=-q_3\left[
<\tilde{\bm p},\bm x>^2
- <\tilde{\bm \alpha},\bm x>^2
- <\tilde{\bm \beta},\bm x>^2
\right]^{1/2},
\eeq
where $ <\tilde{\bm p},\bm x>\equiv \tilde p_0x_0-\tilde p_1 x_1-\tilde
p_2 x_2-\tilde p_3 x_3$, $ <\tilde{\bm \alpha},\bm x>\equiv
\tilde \alpha_0x_0-\tilde \alpha_1 x_1-\tilde
\alpha_2 x_2-\tilde \alpha_3 x_3$, and
$ <\tilde{\bm \beta},\bm x>\equiv
\tilde \beta_0x_0-\tilde \beta_1 x_1-\tilde
\beta_2 x_2-\tilde \beta_3 x_3$.
Here $c$ and $q_3$ are arbitrary nonzero real constants, whereas the rest
of the real parameters have to satisfy condition (\ref{314}) and
\bg
<\tilde{\bm \gamma},\tilde{\bm\gamma}>=-1,\quad
<\tilde{\bm \gamma},\tilde{\bm p}>=
<\tilde{\bm \beta},\tilde{\bm\gamma}>=
<\tilde{\bm \alpha},\tilde{\bm\gamma}>=0,\quad q_1^2+q_2^2=q\neq 0.
\ed
%
%\tilde\gamma_0\tilde\gamma_0-\tilde\gamma_1\tilde\gamma_1
%-\tilde\gamma_2\tilde\gamma_2
%-\tilde \gamma_3\tilde\gamma_3=-1,\quad
%\tilde p_0\tilde\gamma_0-\tilde p_1\tilde\gamma_1
%-\tilde p_2\tilde\gamma_2
%-\tilde p_3\tilde\gamma_3=0\\
%& &\tilde \beta_0\tilde\gamma_0-\tilde\beta_1\tilde\gamma_1
%-\tilde \beta_2\tilde\gamma_2
%-\tilde \beta_3\tilde\gamma_3=
%\tilde \alpha_0\tilde\gamma_0-\tilde\alpha_1\tilde\gamma_1
%-\tilde \alpha_2\tilde\gamma_2
%-\tilde \alpha_3\tilde\gamma_3=0\\
%& & q_1^2+q_2^2=q\neq 0.
%\ed
By the above Ans\"atze the following ODEs are respectively  obtained:
\beg
\label{nop_1}
& & (2c\omega_1+q)\frac{d^2\varphi}{d\omega_1^2}+2c\frac{d\varphi}{d\omega_1}
-\lambda_1\varphi+\lambda_2\varphi^3=0,\\
& &\nonumber\\
\label{nop_2}
& & q_3\omega_2\frac{d^2\varphi}{d\omega_2^2}+2q_3\frac{d\varphi}{d\omega_2}
+\lambda_1\omega_2\varphi-\lambda_2\omega_2\varphi^3=0.
\eeq

Equations (\ref{nop_1}) and (\ref{nop_2}) are not of Painlev\'e type,
which is in agreement with the fact
that
$\omega_1$ and $\omega_2$ do not satisfy the 4-dimensional Bateman
equation $\det B_5^4=0$.

\strut\hfill

A systematic classification of integrable reductions of the above
given multidimensional wave equations, by the use of the Propositions
and Corollary stated here,
will be the subject of a future paper.

\strut\hfill

%\pagebreak

\begin{center}
{\large{\bf References}}
\end{center}
\begin{itemize}
\item[1.]
Ablowitz M.J. and Clarkson P.A.: {\it Solitons, nonlinear evolution
equations and inverse scattering}, Cambridge University Press,
Cambridge, 1991.
\item[2.]
Ablowitz M.J., Ramani A. and Segur H.: 
A connection between nonlinear evolution equations and ordinary
differential equations of P-type. I and II, {\it J. Math. Phys.} {\bf
21} (1980), 715--721; 1006--1015.
\item[3.]
Cariello F. and Tabor M.: Painlev\'e expansion for nonintegrable
evolution equations, {\it Physica} {\bf D39} (1989), 77--94.
\item[4.]
Conte R.: The Painlev\'e approach to nonlinear ordinary differential
equations, in: R. Conte (ed),
{\it The Painlev\'e property, one century later}, Springer-Verlag,
Berlin, 1998. 
\item[5.]
Davis H.T.: {\it Introduction to nonlinear differential and integral
equations}, Dover Publications, New York, 1962.
\item[6.]
Est\'evez P.G., Conde E. and Gordoa P.R.: Unified approach to Miura,
B\"acklund transformation and Darboux transformations for nonlinear
partial differential equations, {\it J. Nonlin. Math. Phys.} {\bf 5} (1998),
82--114.
\item[7.]
Euler N.: Painlev\'e series for $(1+1)$- and $(1+2)$-dimensional
discrete-velocity Boltzmann equations, {\it Research Report} 1997:7,
Dept. of Math.,  Lule\aa\ University of
Technology, ISSN 1400-4003 (21 pages).
\item[8.]
Euler N., Steeb W.-H. and Cyrus K.: Polynomial field theories and
nonintegrability, {\it Physica Scripta} {\bf 41} (1990), 298--301.
\item[9.]
Fairlie D.B.: Integrable systems in higher dimensions,
{\it Prog. of Theor. Phys. Supp.} No. 118 (1995), 309--327.
\item[10.]
Fushchich W.I, Shtelen W.M. and Serov N.I.: {\it Symmetry analysis and
exact solutions of equations of nonlinear mathematical physics},
Kluwer Academic Publishers, Dordrecht, 1993.
\item[11.]
Kovalevskaya S.: Sur le probl\`me de la rotation d'un corps solide
autour d'un
point fixe, {\it Acta Mathematica} {\bf 12} (1889), 177-232.
\item[12.]
Kruskal, M.D., Joshi N. and Halburd R.:
Analytic and asymptotic methods for nonlinear singularity analysis: a
review and extension of tests for the Painlev\'e property, in: {\it
Integrability of nonlinear systems} Volume 495 of {\it Lecture notes
in Physics}, 171--205, Springer-Verlag, Heidelberg, 1997.
\item[13.]
McLeod J.B. and Olver P.J.: The connection between partial
differential equations soluble by inverse scattering and ordinary
differential equations of Painlev\'e type, {\it SIAM J. Math. Anal.} {\bf
14} (1983), 488--506.
\item[14.]
Mitropolskii, Yu.A. and Shul'ga M.V.: Asymptotic solutions of a
multidimensional nonlinear wave equation, {\it Soviet Math. Dokl.} {\bf
36} (1987), 23--26.
\item[15.]
Musette, M.: Painlev\'e analysis for nonlinear partial differential
equations, in: R. Conte (ed),
{\it The Painlev\'e property, one century later}, Springer-Verlag,
Berlin, 1998.
\item[16.]
Newell A.C., Tabor M. and Zeng Y.B.: A unified approach to
Painlev\'e expansions, {\it Physica} {\bf 29D} (1987), 1--68.
\item[17.]
Painlev\'e, P.: Sur les \'equations diff\'erentielles du
premier ordre, {\it C.R. Acad. Sc. Paris} {\bf 107} (1888),
221--224, 320--323, 724--726.
\item[18.] Painlev\'e, P.: M\'emoire sur les \'equations
diff\'erentialles dont l'integrale g\'en\'erale est uniforme,
{\it Bull. Soc. Math. France} {\bf 28} (1900), 201-261.
\item[19.]
Painlev\'e, P.: Sur les \'equations diff\'erentielles du second ordre
\`a
points critiques fix\`es, 
{\it C.R. Acad. Sc. Paris} {\bf 143} (1906), 1111-1117.
\item[20.]
Steeb W.-H. and Euler N.: Nonlinear evolution equations and
Painlev\'e test, World Scientific, Singapore, 1988.
\item[21.]
Ward R.S.: The Painlev\'e property for self-dual gauge-field
equations, {\it Phys. Lett.} {\bf 102A} (1984), 279--282.
\item[22.]
Webb G.M. and Zank G.P.: Painlev\'e analysis of the
three-dimensional Burgers' equation, {\it Phys. Lett.}  {\bf A150}
(1990), 14--22.
\item[23.]
Webb G.M. and Zank G.P.: On the Painlev\'e analysis of the
two-dimesional Burgers' equation, {\it Nonl. Anal. Theory Meth. Appl.}
{\bf 19} (1992), 167--176.
\item[24.]
Weiss J.: The Painlev\'e property for Partial differential
equations II: B\"acklund transformations, Lax pairs, and the
Schwarzian derivative, {\it J. Math. Phys.} {\bf 24} (1983), 1405--1413.
\item[25.]
Weiss J.: The sine-Gordon equation: Complete and partial
integrability, {\it J. Math. Phys.} (1984) {\bf 25}.
\item[26.]
Weiss J., Tabor M. and Carnevale G.: The Painlev\'e property
for partial differential equations, {\it J. Math. Phys.} {\bf 24}
(1983),
522--526.
\end{itemize}

\end{document}